\documentclass[epj]{webofc}
\usepackage[varg]{txfonts}   
\usepackage{color}
\usepackage{lipsum}
\usepackage{wrapfig}
\usepackage{caption}
\usepackage{subcaption}
\woctitle{Mathematical Modeling and Computational Physics 2015}

\usepackage[english]{babel}
\begin{document}
\title{An Algorithm for the Simulations of the Magnetized\\Neutron Star Cooling}
%
%

\author{H.~Grigorian\inst{1,2}\fnsep\thanks{\email{hovikgrigorian@gmail.com}}
        \and
        A.~Ayriyan\inst{1}
        \and
        E.~Chubarian\inst{2}
        \and
        A.~Piloyan\inst{3}
        \and
        M.~Rafayelyan\inst{4}
}

\institute{
		Joint Institute for Nuclear Research, Joliot-Curie~6, 141980 Dubna, Moscow Region, Russia
\and
		Yerevan State University, Alek Manyukyan~1, 0025 Yerevan, Republic of Armenia
\and
		Yerevan Physics Institute, A. Alikhanian Brothers~2, 0036 Yerevan, Republic of Armenia
\and
		University of Bordeaux, 351 Cours de la Lib\'{e}ration, F-33400 Talence, France
          }

\abstract{%
The model and algorithm for the cooling of the magnetized neutron stars are presented. The cooling evolution described by system of parabolic partial differential equations with non-linear coefficients is solved using Alternating Direction Implicit method. The difference scheme and the preliminary results of simulations are presented.
}
\maketitle

\section{Introduction}
\label{intro}

\begin{wrapfigure}{r}{0.45\textwidth}
	\vspace{-8mm}
	\begin{center}
		\includegraphics[width=0.4\textwidth]{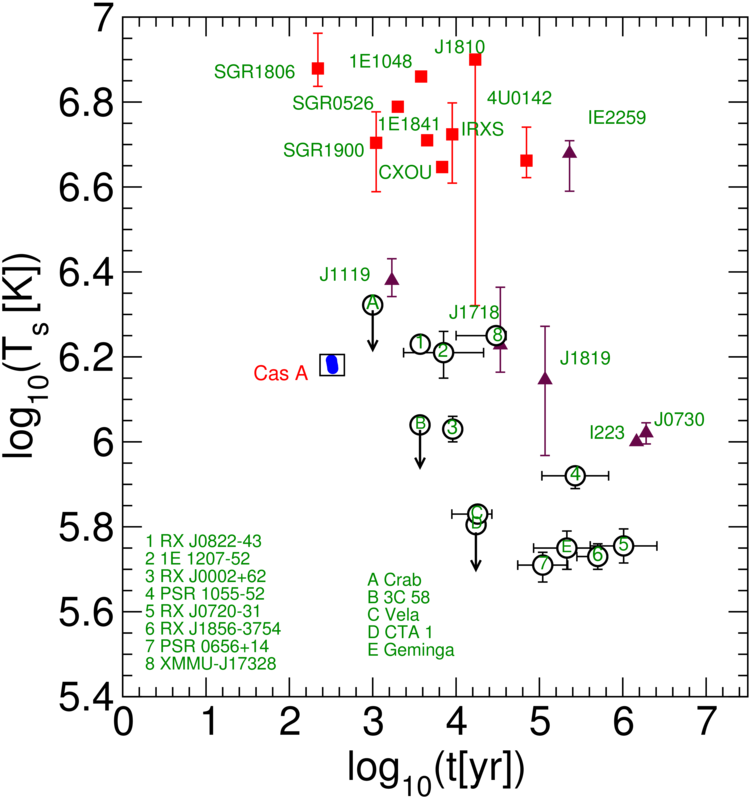}
	\end{center}
	\vspace{-6mm}
	\caption{Observational data of age and temperature of compact stars. Squares -- AXPs and SGRs neutron stars with magnetic field from $10^{14}~\textrm{G}$ up to $10^{15}~\textrm{G}$. Triangles -- radio-quiet stars with field around $10^{13}~\textrm{G}$. Circles -- radio-pulsars with field $\le 10^{12}~\textrm{G}$.}
	\vspace{-7mm}
	\label{fig:obs_data}
\end{wrapfigure}

Modelling of magnetized neutron stars cooling process is an interesting subject for the
investigations of the internal structure of such objects. The~main aspect is the investigation
of the nuclear matter properties in extremely high densities. The~magnetic field of compact object
when it is more than $10^{14}~\textrm{G}$ has a significant influence on the heat transport inside the star
and can give observational effects on the surface temperatures.
The model of cooling evolution is given by a system of parabolic partial differential equations
with non-linear temperature depending thermal coefficients and cooling (via neutrinos) and heating (due to magnet field decay) sources.

Due to the axial symmetry of the field and spherical symmetry of the matter distribution inside, the problem can be realized with the choice of spherical coordinates in spatial 2D. For the solution we choose the Alternating Direction Implicit (ADI) method \cite{samarski_1995, yanenko_1967}. In the difference scheme we use non-constant spatial steps and self-correcting time step. We investigate the special boundary conditions in the center and on the surface of the star configuration. At the current stage of the study we are going to present some preliminary results of the cooling simulations to demonstrate the efficiency of our 2D cooling algorithm.

The nowadays observational data of the magnetized neutron star~(Fig.~\ref{fig:obs_data}, the table of data with the corresponding references see in~\citep{Aguilera_2008_AJ}) could not be explained by~1D cooling simulations neglecting magnetic field inside. Inclusion of the magnetic field is necessary but even not sufficient yet \cite{Aguilera_2008_AJ, Aguilera_2008_AA}. In the current work we focus on the algorithm of 2D simulation of cooling of magnetized neutron star.

\section{Equations for Thermal Evolution of the Magnetized Compact Star}

The neutron star is bounded by gravitational field, which in the approximation of slow rotation of the stars can be described by a metric tensor of the space-time manifold~\cite{Misner1973}
\begin{equation}
ds^{2}=-e^{2\Phi}dt^{2}+e^{2\Lambda}dr^{2}+r^{2}d\Omega^{2}.\label{eq:metrics}
\end{equation}

The compact star configuration is constructed with the use of equation of state (EoS) of the stellar matter based on knowledge of nuclear matter~\cite{Weber_1996}.
The coefficients of the metric tensor as well as characteristics of the matter distribution are self-consistent solutions of the  Einstein equation (specially for this case TOV \cite{Misner1973, Weber_1996}) where the temperature effect on internal structure could be neglected because of high density inside the star.

On the other hand the thermal evolution of the star can be described with the use of energy balance equations, which has parabolic equation form:
\begin{eqnarray}
\label{eq:balance}
c_{v}e^{\Phi}\dfrac{\partial T}{\partial t} & +\vec{\nabla}\cdot(e^{2\Phi}\vec{F}) & =e^{2\Phi}Q,
\end{eqnarray}
where $c_{v}$ is the specific heat per unit volume and $Q$ is the energy loss/gain by neutrino emission, Joule heating, accretion heating, etc. The vector $\vec{F}$ is corresponding to the heat flux. 

The thermal properties of the stellar matter are investigated in frame of different choices of nuclear matter cooling scenarios~\cite{Grigorian_2004_AA, Grigorian_2005_AA, Page_2004_AJS, Page_2009_AJ, Grigorian_2005_PRC}.

The flux is created due to the temperature gradient and tends to equilibrate the temperature 
\begin{eqnarray}
\label{eq:flux}
\nonumber \vec{F} & = & -e^{-\Phi}\hat{\kappa}\cdot\vec{\nabla}(e^{\Phi}T).
\end{eqnarray}
Here $\hat{\kappa}$ is the thermal conductivity, which is a tensor in the case of enough strong magnetic field of the star.

Hereafter we will use the notations $\tilde{T}\equiv e^{\Phi}T$ and $\tilde{F}\equiv e^{2\Phi}F$. The flux components (in spherical coordinates) in the case of dipole magnetic field are: 
\begin{eqnarray}
\label{eq:hhh}
\nonumber \tilde{F}_{r} = -e^{\Phi}\left(\kappa_{rr}e^{-\Lambda}\partial_{r}\tilde{T}+\dfrac{\kappa_{r\theta}}{r}\partial_{\theta}\tilde{T}\right), &
\tilde{F}_{\theta} = -e^{\Phi}\left(\kappa_{\theta r}e^{-\Lambda}\partial_{r}\tilde{T}+\dfrac{\kappa_{\theta\theta}}{r}\partial_{\theta}\tilde{T}\right),
\end{eqnarray}
when the $\phi$-component is zero due to the axial symmetry.

Influence of the magnetic field, which effects the conductivity, leads to anisotropy along and orthogonal to the field. The relation between conductivity components is defined in terms of the magnetization parameter $\omega_{B}\tau$:
\begin{equation}
\dfrac{\kappa_{\parallel}}{\kappa_{\perp}}=1+(\omega_{B}\tau)^{2},
\end{equation}
where $\tau$ is the particle relaxation time \cite{Urpin_Yakovlev_1980}, and $\omega_{B}$ is the classical gyrofrequency corresponding to the magnetic field strength $B$:
\begin{equation}
\omega_{B} = \dfrac{eB}{mc}.
\end{equation}
$m$ is the effective mass and $e$ is a charge of a particle involved with magnetic field, here $c$ is the speed of light.

With the use of unit vector $\vec{b}$ of magnetic dipole the flux can be expressed in the form: 
\begin{equation}
\vec{F} = -e^{\Phi}\kappa_{\perp}\left[\vec{\nabla}\tilde{T}+\left(\omega_{B}\tau\right)^{2}\left(\vec{b}\cdot\vec{\nabla}\tilde{T}\right)\cdot\vec{b}+\omega_{B}\tau\left(\vec{b}\times\vec{\nabla}\tilde{T}\right)\right].
\end{equation}

When the magnetic field is parallel to the axis $z$ one has $b=(\cos\theta,-\sin\theta,0)$. So for the components of the conductivity we have:
\begin{eqnarray}
\kappa_{rr} & = & \dfrac{1+(\omega_{B}\tau)^{2}\xi^{2}}{1+(\omega_{B}\tau)^{2}}\kappa_{\parallel}\,,\nonumber\\
\kappa_{\theta r} & = & \kappa_{r\theta} =\xi\sqrt{1-\xi^2}(\kappa_{\parallel}-\kappa_{\bot})\,,\nonumber\\
\kappa_{\theta\theta} & = & \dfrac{1+(\omega_{B}\tau)^{2}(1-\xi^{2})}{1+(\omega_{B}\tau)^{2}}\kappa_{\parallel}\,.\nonumber
\end{eqnarray}
For simplicity we will use the following notations: $\xi\equiv \cos(\theta)$, $\tilde{Q}\equiv e^{2\Phi}Q$. 

The introduction of the new variables $\xi$ and the accumulated mass $m=4\pi\int\rho r^{2}dr$ (here $\rho$ is the energy density of the matter) instead of $\theta$ and $r$ leads the thermal evolution equation~(\ref{eq:balance}) to the following form
\begin{equation}
\nonumber \dfrac{\partial u}{\partial t}  =D_{1}\dfrac{\partial}{\partial m}\left(B_{1m}\dfrac{\partial u}{\partial m}+B_{1\xi}\dfrac{\partial u}{\partial\xi}\right)+D_{2}\dfrac{\partial}{\partial\xi}\left(B_{2m}\dfrac{\partial u}{\partial m}+B_{2\xi}\dfrac{\partial u}{\partial\xi}\right) + Q_{B}.
\end{equation}
The sought-for function is $u=\log(\tilde{T})$ and the coefficients are
\begin{eqnarray}
B_{1m} \equiv 4\pi r^{4}\rho e^{-\Lambda}(1+(\omega_{B}\tau)\xi^{2})k_{\bot}\tilde{T}e^{\Phi}, &
B_{2m} \equiv 4\pi r\rho e^{-\Lambda}\xi\left(1-\xi^{2}\right(\omega_{B}\tau)^{2}k_{\bot}\tilde{T}e^{\Phi},\nonumber\\
B_{1\xi} \equiv r\xi\left(1-\xi^{2}\right)(\omega_{B}\tau)^{2}k_{\bot}\tilde{T}e^{\Phi}, &
B_{2\xi} \equiv \left(1+(\omega_{B}\tau)^{2}\left(1-\xi^{2}\right)\right)\left(1-\xi^{2}\right)k_{\bot}
\tilde{T}e^{\Phi}/r^{2},\nonumber
\end{eqnarray}
and $D_{1}\equiv4\pi D_{2}\rho e^{-\Lambda}$, $D_{2}\equiv(c_{v}\widetilde{T})^{-1}$, $Q_{B}\equiv D_{2}\tilde{Q}$.

\section{2D Scheme for Mixed Implicit and Explicit Methods}

For the numerical calculation we use the most convenient from stability point of view ADI method~\cite{samarski_1995, yanenko_1967} and discretize the direction $m$ taking into account the energy distribution $\rho(r)$. This function is the solution of a single compact star for given central density or given total mass. So, corresponding to each given distance from the center $r_{j}$ we have points of grid $m_{j}$. The mass for the given $j$ is the same for all angles, therefore $\xi$ could be considered as an independent coordinate.

The steps of discretization are $\triangle m_{j\pm1/2}=\pm\left(m_{j\pm1}-m_{j}\right)$ and $\triangle m_{j}=(1/2)\left(\Delta m_{j+1/2}+\Delta m_{j-1/2}\right)=(1/2)\left(m_{j+1}-m_{j-1}\right)$. Similar notations we use for the angle $\xi$ as well as for any function $F(r,\theta)$: $F_{i,j\pm1/2} = (1/2)\left(F_{i,j\pm1}+F_{i,j}\right)$.

So for the first order partial derivatives one has
\begin{eqnarray}
\left(\dfrac{\partial u}{\partial m}\right)_{i,j\pm1/2}=\pm\dfrac{u_{i,j\pm1}-u_{ij}}{\triangle m_{j\pm1/2}}, &  &
\left(\dfrac{\partial u}{\partial\xi}\right)_{i\pm1/2,j}=\pm\dfrac{u_{i\pm1,j}-u_{ij}}{\triangle\xi_{i\pm1/2}},\nonumber
\end{eqnarray}
and for the second order onces we introduced $R$ and $S$ operators correspondingly for the same and mixed variables: 
\begin{equation*}
\left[ R^{(1)}u\right]_{ij}\equiv-D_{1}\dfrac{\partial}{\partial m}\left(B_{1m}\dfrac{\partial u}{\partial m}\right)_{i,j}=-\dfrac{D_{1i,j}}{\triangle m_{j}}\left(B_{1m,i,j+1/2}\dfrac{u_{i,j+1}-u_{i,j}}{\triangle m_{j+1/2}}-B_{1m,i,j-1/2}\dfrac{u_{ij}-u_{ij-1}}{\triangle m_{j-1/2}}\right),
\end{equation*}
\begin{equation*}
\left[ R^{(2)}u\right]_{ij}\equiv-D_{2}\dfrac{\partial}{\partial\xi}\left(B_{2\xi}\dfrac{\partial u}{\partial\xi}\right)_{i,j}=-\dfrac{D_{2i,j}}{\triangle\xi_{i}}\left(B_{2\xi,i+1/2,,j}\dfrac{u_{i+1,j}-u_{i,j}}{\triangle\xi_{i+1/2}}-B_{2\xi i-1/2,j}\dfrac{u_{ij}-u_{i-1,j}}{\triangle\xi_{i-1/2}}\right),
\end{equation*}
and
\begin{equation*}
\left[ S^{(1)}u\right]_{ij}=D_{1}\dfrac{\partial}{\partial m}\left(B_{1\xi}\dfrac{\partial u}{\partial\xi}\right)_{ij}=\dfrac{D_{1i,j}}{2\triangle m_{j}}\left(B_{1\xi,i,j+1}\dfrac{u_{i+1,j+1}-u_{i-1,j+1}}{2\triangle\xi_{i}}-B_{1\xi,i,j-1}\dfrac{u_{i+1j-1}-u_{i-1,j-1}}{2\triangle\xi_{i}}\right),
\end{equation*}
\begin{equation*}
\left[ S^{(2)}u\right]_{ij}=D_{2}\dfrac{\partial}{\partial\xi}\left(B_{2m}\dfrac{\partial u}{\partial m}\right)_{i,j}=\dfrac{D_{2i,j}}{2\triangle\xi_{i}}\left(B_{2m,i+1,j}\dfrac{u_{i+1,j+1}-u_{i+1,j-1}}{2\triangle m_{j}}-B_{2m,i-1,j}\dfrac{u_{i-1j+1}-u_{i-1,j-1}}{2\triangle m_{j}}\right).
\end{equation*}
Finally after discretization of the time with the steps $\Delta t_{n}$ the differential equation of the problem can be written as:
\begin{equation*}
\dfrac{u^{n+1}-u^{n}}{\triangle t_{n}}+\sigma R^{(1)}u^{n+1}+(\sigma-1)R^{(2)}u^{n+1}=(\sigma-1)R^{(1)}u^{n}-\sigma R^{(2)}u^{n}+S^{(1)}u^{n}+S^{(2)}u^{n}+Q_{B}.
\end{equation*}

The introduced parameter $\sigma$ stands for ADI method: when $\sigma=1$ we apply implicit method in $m$ direction and when $\sigma=0$ in $\xi$ direction.

The tri-diagonal elements and right hand side of the algebraic equations can be written in the following form
\begin{eqnarray}
\label{eq:tri-diag}
X^{[i][j]} & = & \sigma\dfrac{K_{i,j+1/2}^{(1)}}{\left(\Delta m_{j}\right)^{2}}+(\sigma-1)\dfrac{K_{i+1/2,j}^{(2)}}{\left(\Delta\xi_{i}\right)^{2}},\nonumber\\
Z^{[i][j]} & = & \sigma\dfrac{K_{i,j-1/2}^{(1)}}{\left(\Delta m_{j}\right)^{2}}+(\sigma-1)\dfrac{K_{i-1/2,j}^{(2)}}{\left(\Delta\xi_{i}\right)^{2}},\\
Y^{[i][j]} & = & 1/\triangle t_{n}-X^{[i][j]}-Z^{[i][j]},\nonumber\\
S^{[i][j]} & = & u_{ij}^{n}/\triangle t_{n} + (\sigma-1)\left[ R^{(1)}u^{n}\right]_{ij}-\sigma\left[ R^{(2)}u^{n}\right]_{ij}+\left[ S^{(1)}u^{n}\right]_{ij}+\left[ S^{(2)}u^{n}\right]_{ij}+\left[ Q_{B}\right]_{ij},\nonumber
\end{eqnarray}
where we use the following notations: $X^{[i][j]}$ and $Z^{[i][j]}$ are upper and lower diagonal vectors correspondingly,  $Y^{[i][j]}$ is the main diagonal vector. For solving this tri-diagonal algebraic system the Thomas algorithm is used \cite{thomas_1949,press_2007}.

\begin{wrapfigure}{r}{0.45\textwidth}
	\vspace{-8mm}
	\begin{center}
		\includegraphics[width=0.45\textwidth]{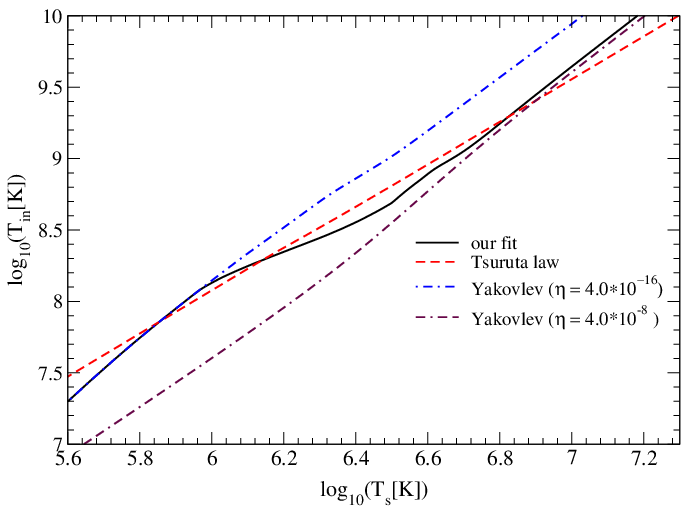}
	\end{center}
	\vspace{-5mm}
	\caption{Relation between the mantle temperature $T_{\rm m}$ and the surface temperature $T_{\rm s}$}
	\vspace{-7mm}
	\label{fig:surface_temp}
\end{wrapfigure}

In the formula~(\ref{eq:tri-diag}) the $K_{i,j}^{(1)}$ and $K_{i,j}^{(2)}$ are the centred mean values of
\begin{eqnarray*}
\nonumber K_{i,j\pm1/2}^{(1)} = D_{1i,j}B_{1mi,j\pm1/2}\dfrac{\Delta m_{j}}{\Delta m_{j\pm1/2}},\\
K_{i\pm1/2,j}^{(2)} = D_{2i,j}B_{2\xi i\pm1/2,j}\dfrac{\Delta\xi_{j}}{\Delta\xi_{j\pm1/2}}.
\end{eqnarray*}

To complete the tri-diagonal algebraic system we put the set of boundary conditions to the physical requirements, namely in the center and on the surface of the neutron star~\cite{Yakovlev_2004_AA}, as follows. All~fluxes of heat from all directions have to vanish at the center point, which leads to $X^{[i][0]}=0$. Therefore, the temperature at the center point could be set as mean value of temperatures in very close area of the center.

At the surface we assume that the flux of the energy towards the radial direction has to be set equal to the flux of energy of the photon emission:
\begin{equation*}
L_{m}(i,J)=2\pi R^{2} \sigma \tilde{T}_{s}^{4}(i) \Delta\xi_i,
\end{equation*}
where $J$ is the maximal value of index $j$, $\sigma$ is the Boltzmann constant and $T_{s}$ is the surface temperature, which is connected to the $T_m$ -- mantle temperature under the core (see Fig.~\ref{fig:surface_temp}):
\begin{equation*}
u_{i,J} = \log{\tilde{T}_m(i)}.
\end{equation*}
For details one can see~\cite{Grigorian_2004_AA}.

From the symmetry reasons the tangential fluxes on the polar and equatorial planes also are zeros $L_{\xi}(i=0,j)=L_{\xi}(i=I,j)=0$:
\begin{equation*}
L_{\xi}(i,j)=\left(B_{1\xi}\dfrac{\partial u}{\partial\xi}\right)_{i,j}=B_{1\xi,i,j}\dfrac{u_{i+1,j}-u_{i-1,j}}{2\Delta\xi_{i}}.
\end{equation*}

\section{Some Features of Simulations and Conclusions}
\label{results}
To demonstrate the work of our algorithm we provide simulations with model coefficients, which are taking into the account the main features of physical properties of thermal coefficients. Particularly, we use cooling mechanism of neutrino emission -- so called modified URCA process (see for details~\cite{Grigorian_2010}). We also vary the magnetic field, however we don't include the Joule heating from the decay of magnetic field.

\begin{figure}[h!]
    \centering
    \begin{subfigure}[b]{0.48\textwidth}
        \centering
        \includegraphics[width=\textwidth]{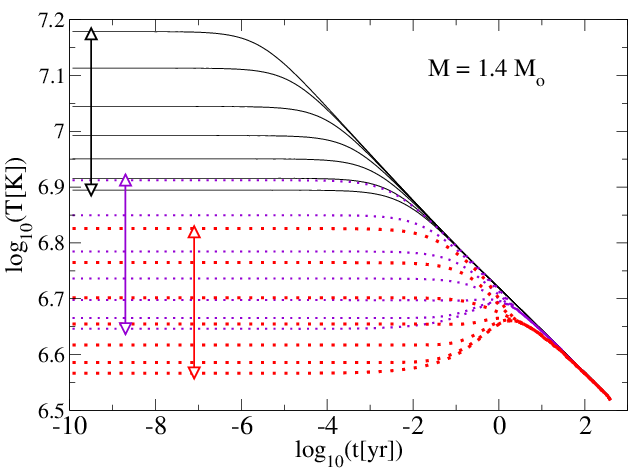}
        \caption{Surface temperature evolution of neutron star with mass $m={1.4}_{\mathrm{sun}}$. Each color corresponds to cooling curves for different azimuthal angles of one object.}
        \label{fig:temp_evol}
    \end{subfigure}
    \hfill
    \begin{subfigure}[b]{0.48\textwidth}
        \centering
        \includegraphics[width=\textwidth]{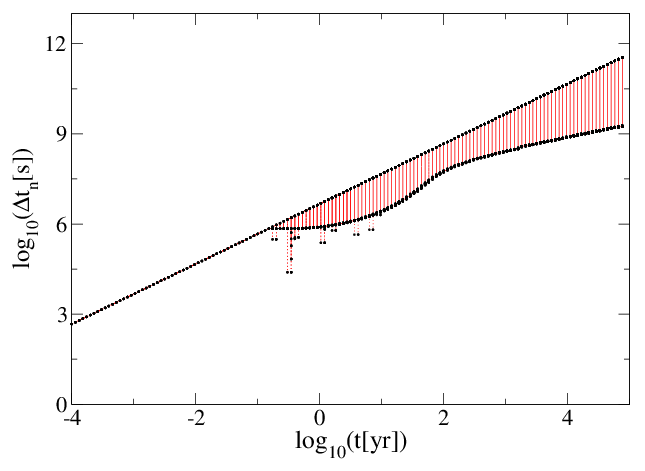}
        \caption{The choice of the time-step during the numerical simulation. \\}
        \label{fig:time_step}
    \end{subfigure}
    \label{fig:results}
    \vspace{-5mm}
    \caption{Simulation results of the coolings of magnetars}
    \vspace{-5mm}
\end{figure}

As it can be seen from the Fig.~\ref{fig:temp_evol} due to absence of additional anisotropic heating sources 
the temperature distribution in all directions becomes isotropic after $50~\mathrm{years}$ of evolution. And we can notice as well that this property is independent from the value of initial anisotropic distribution of the temperature.

One can notice that if temperature in some direction initially less then the temperature at the beginning of the isotropic era it heats up during the evolution (see red lines on Fig.~\ref{fig:temp_evol}). It means that heat conducting process dominates over neutrino cooling.

The next, we note that this period $50-100~\mathrm{years}$ are short enough to be able to compare this results with observations, because all data we have as it shown in Fig.~\ref{fig:obs_data} are for objects older than at least $500~\mathrm{years}$. On the other hand one can see that the temperature values at the beginning of the isotropic era are around the same what are the observational temperatures of already cold magnetars.

It means that long stay of the heat inside the star for such longer time could not be just the effect of anisotropy due to the magnetization. It is even hard to argue that the observed temperatures of red point in Fig.~\ref{fig:obs_data} are estimated only for small area around the poles of the stars.

On the~Fig.~\ref{fig:time_step} we demonstrate the automatic choice of the time-step of the numerical algorithm. The~top line on this figure is the predicted time-step, the bottom curve is the modified time-step to provide stability of the algorithm. The automatic choice of time-step is needed mainly due to non-trivial mass distribution inside the star.

\begin{acknowledgement}
The study is partially supported by RFBR according to the scientific project No.~14-01-00628. Authors thank Dr.~Edik Ayryan (LIT, JINR) for fruitful discussions and supporting the participation of HG on MMCP'2015.
HG, AA, MR acknowledge Ter-Antonyan--Smorodinskii Program for supporting exchange between JINR Dubna and Armenian Universities and Institutes.
HG, EC, AP and MR thank Ministry of Education and Science of Armenia for the  support according to the scientific project of State Committee of Science, grant No.~H-11-1c259.
\end{acknowledgement}

%
%
%

\end{document}